\documentclass[letterpaper,10pt,conference]{ieeeconf}
\IEEEoverridecommandlockouts
\usepackage{flushend}

\usepackage[compress]{cite}
\usepackage{amsmath,amssymb,amsfonts}
\usepackage{graphicx}
\usepackage{textcomp}
\usepackage{booktabs,tabularx,makecell,array,multirow}

\usepackage[hidelinks]{hyperref}
\usepackage{enumerate}

\usepackage{mathtools}
\newcommand{\defeq}{\coloneqq}

\usepackage[capitalize,nameinlink]{cleveref}
\usepackage[font=footnotesize]{caption}

\newcommand{\bmat}[1]{\begin{bmatrix}#1\end{bmatrix}}

\newcommand{\squeezemat}{\addtolength{\arraycolsep}{-1mm}}

\graphicspath{{./figs/}}

\newcommand{\ee}{\mathbb{E}}
\newcommand{\prob}{\mathbb{P}}
\newcommand{\dd}{\,\mathrm{d}}
\newcommand{\grad}[1]{\nabla #1}
\newcommand{\hess}[1]{\nabla^2 #1}

\newcommand{\R}{\mathbb{R}}
\newcommand{\N}{\mathcal{N}}

\newcolumntype{Y}{>{\centering\arraybackslash}X}

\DeclareMathOperator{\diag}{diag}
\DeclareMathOperator*{\argmax}{argmax}

\newtheorem{lemma}{Lemma}
\newtheorem{definition}{Definition}
\newtheorem{remark}{Remark}

\newtheorem{proposition}{Proposition}
\newtheorem{corollary}{Corollary}

\renewcommand{\epsilon}{\varepsilon}

\title{\LARGE \bf
State Estimation for Linear Systems with Non-Gaussian\\Measurement Noise via Dynamic Programming
}

\author{Mohammad Hussein Yoosefian Nooshabadi and Laurent Lessard%
\thanks{Authors are with the Department of Mechanical and Industrial Engineering, Northeastern University, Boston, MA 02115, USA. Email:\newline
{\tt\{yoosefiannooshabad.m,\,l.lessard\}@northeastern.edu}}
}

\begin{document}

\maketitle

\begin{abstract}
We propose a new recursive estimator for linear dynamical systems under Gaussian process noise and non-Gaussian measurement noise. Specifically, we develop an approximate \emph{maximum a posteriori} (MAP) estimator using dynamic programming and tools from convex analysis. Our approach does not rely on restrictive noise assumptions and employs a Bellman-like update instead of a Bayesian update. Our proposed estimator is computationally efficient, with only modest overhead compared to a standard Kalman filter. Simulations demonstrate that our estimator achieves lower root mean squared error (RMSE) than the Kalman filter and has comparable performance to state-of-the-art estimators, while requiring significantly less computational power.
\end{abstract}

\section{Introduction}\label{sec: intro}

We consider state estimation for discrete-time linear systems driven by non-Gaussian noise, in the standard form
\begin{subequations} \label{eq: general dyn sys}
  \begin{align}
    x_{t} & = Ax_{t-1} + w_{t},\\
    y_t & = Cx_t + v_t,
\end{align}  
\end{subequations}
where $A$ and $C$ are known, and the distributions of the process noise $w_t$, measurement noise $v_t$, and initial state $x_0$ are specified and not necessarily Gaussian.
We make the standard assumptions that $w_t$ and $v_t$ are independent of each other, of $x_t$, and across time.

In this setting, the exact posterior distribution given past measurements is provided by \emph{Bayesian filtering}, which is a recursive formula for updating the conditional state distribution when a new measurement $y_t$ arrives
\begin{multline}\label{eq:bayesian_update}
\prob(x_t\mid y_{0:t}) \\
\propto
\prob(y_t\mid x_t)\int \prob(x_t\mid x_{t-1})\,\prob(x_{t-1}\mid y_{0:t-1})\dd x_{t-1}.
\end{multline}
The estimator that minimizes the mean squared error (MSE), is given by the conditional expectation
\begin{equation}\label{eq: mmse estimator}
\hat x_t = \ee( x_t \mid y_{0:t}) = \int x_t\, \prob(x_t \mid y_{0:t})\dd x_t.
\end{equation}
When the noise and the initial state are Gaussian, the integral in \eqref{eq:bayesian_update} has a closed-form solution, and the estimator \eqref{eq: mmse estimator} can be found recursively via the celebrated Kalman filter (KF).

Despite the widespread applications of the KF \cite{han2020blind, abdollahi2024improved, taassori2024enhancing}, its Gaussianity assumption does not hold in many practical scenarios. Notable examples include underwater communication \cite{wang2020novel}, power systems \cite{sarfi2019decentralized} and magnetic resonance imaging (MRI) \cite{st2018automatic}, where Gaussian mixture or various heavy-tailed distributions have been reported. In the non-Gaussian setting, the integral in \eqref{eq:bayesian_update} typically does not have a closed-form solution, so one must resort to approximation techniques or alternatively, use estimators that do not directly deal with the integration \eqref{eq:bayesian_update}. This leads to two general classes of estimators, which we briefly survey below. 

\subsection{Integration-based Estimators}\label{sec: posterior-based methods}

Directly approximating the integral in \eqref{eq:bayesian_update} yields a trade-off between computational effort and approximation error. 
At one extreme, we have \emph{sequential Monte Carlo} approaches, such as bootstrap filtering \cite{gordon1993novel} and particle filtering \cite{elfring2021particle}, where distributions are represented by a set of random samples. These approaches are computationally intensive and also applicable when the dynamics are nonlinear.
At the other extreme, the simplest approach, which we call the \emph{standard KF}, is to replace non-Gaussian noise distributions with Gaussian distributions of matching mean and variance and then apply a standard KF. Despite being computationally attractive, this approach may lead to severely degraded performance when the noise is heavy-tailed or bimodal.

Many other schemes have been proposed that balance computational load and approximation error. In \cite{sorenson1971recursive} the noise and posterior distributions are approximated as a sum of Gaussians, yielding the \emph{Gaussian sum filter}. Despite favorable performance compared to the standard KF, the number of Gaussians required to represent the posterior grows exponentially with the time horizon. The work \cite{bilik2010mmse} proposed a more scalable version of GSF. An alternative approach, proposed by Masreliez \cite{masreliez1975approximate}, approximates the Bayesian update using a \emph{score function}; the gradient of $\log \prob(y_t\mid y_{0:t-1})$. This method outperforms the standard KF, but requires evaluating a difficult integral similar to \eqref{eq:bayesian_update} at each timestep. The work \cite{wu1989kalman} proposed an efficient implementation of the Masreliez estimator via polynomial approximation.\looseness=-1

\subsection{Optimization-based Estimators}\label{sec: non posterior-based methods}

Alternatively, one may forego the Bayesian update \eqref{eq:bayesian_update}, and instead formulate an optimization problem that is solved whenever a new measurement arrives. For such estimators, conditional distributions are never computed or approximated; rather, robustness to non-Gaussianity is enforced by explicitly minimizing an auxiliary cost function. In \cite{chen2017maximum}, a \emph{maximum correntropy Kalman filter} (MCKF) is proposed, which has shown superior performance compared to the standard KF in heavy-tailed noise environments. In \cite{he2023generalized} the \emph{entropy} of the estimation error is minimized instead, which can outperform the MCKF in certain noise environments. The authors in  \cite{kircher2021optimization} develop optimization-based estimators that are resilient to bounded process noise and impulsive measurement noise. Finally, \cite{gultekin2017nonlinear} minimizes the Kullback--Liebler divergence of the estimation error. Although these methods, when tuned properly, offer great performance, they are usually designed for specific measurement noise distributions. Furthermore, since these methods involve auxiliary optimization at each timestep, they tend to be computationally more demanding than the standard KF.

In this paper, we propose a new estimator inspired by \emph{maximum a posteriori} (MAP) estimation.  As we detail in \cref{sec: main results}, the MAP framework yields a Bellman-style recursion rather than Bayesian recursion \eqref{eq:bayesian_update}. Although this Bellman recursion does not have a closed-form solution, under proper approximation of the value function, it reduces to algebraic equations, which we use to develop a nonlinear estimator for the case with Gaussian process noise and non-Gaussian measurement noise. Simulations show that our proposed estimator achieves lower RMSE than the Kalman filter and matches the performance of state-of-the-art alternatives at a fraction of their computational cost.

The rest of the paper is organized as follows. We present preliminaries and assumptions in \cref{sec: problem setup}, our main results in \cref{sec: main results}, discussion in \cref{sec: discussion}, and numerical examples in \cref{sec: numerical examples}. Finally,  we conclude and discuss future research directions in \cref{sec: conclusion and future works}.

\section{Preliminaries and Assumptions} \label{sec: problem setup}

\subsection{Notation}
The notation $\prob$ and $\ee$ denote probability density function (PDF) and expectation of a random variable or vector.
For a square symmetric real matrix $M$, we write $M \succ 0$ to mean that $M$ is positive definite. We write $I_n\in\R^{n\times n}$ to denote the identity matrix. We write $\N(\mu, \Sigma)$ to denote a (multivariate) Gaussian distribution with mean $\mu$ and covariance $\Sigma \succ 0$. 
Also, $\ell(\cdot) \defeq \log \prob(\cdot)$ and $\ell(\,\cdot\mid\cdot\,) \defeq \log \prob(\,\cdot\mid\cdot\,)$ denote marginal and conditional log-likelihood, respectively.

For the rest of the paper, we use ``$\star$'' to denote submatrices that can be inferred from symmetry and ``$\dots$'' to denote entries omitted because they are irrelevant to the discussion.

\subsection{Assumptions}\label{assumptions}
We consider the dynamical system \eqref{eq: general dyn sys}
where $x_t \in \R^n$, $y_t \in \R^d$, $w_{t} \in \R^n$, and $v_t \in \R^d$ are the state vector, measured output, process noise, and measurement noise, respectively, and $A$ and $C$ are known. 
For convenience, we reparameterize
$\prob(w_{t}) = e^{-q(w_{t})}$ and $\prob(v_t) = e^{-r(v_t)}$, where $q:\R^n\to\R$ and $r:\R^d\to\R$ are continuously differentiable. We also make the mild technical assumption that the mode exists and is unique for all distributions.
These assumptions encompass many common distributions, such as Gaussian mixtures, Cauchy, and skewed normal. 

\subsection{The Kalman Filter}\label{subsec: KF}
Consider the dynamical system \eqref{eq: general dyn sys}. If the initial state and the noise inputs are Gaussian, $x_0 \sim \N(\mu_{0|-1},P_{0|-1})$, $w_{t} \sim \N(\mu_w, \Sigma_w)$ and $v_t \sim \N(\mu_v, \Sigma_v)$, then the state has a Gaussian posterior $(x_t \mid y_0,\dots,y_t) \sim \N(\mu_t,P_t)$, which can be computed recursively via the celebrated Kalman filter (KF). The two-step KF can be written as follows \cite[\S 6.2]{simon2006optimal}.
\begin{itemize}
    \item \emph{Time update:}
    \begin{subequations}\label{eq: KF time update}
    \begin{align}
        P_{t|t-1} & = \Sigma_w + A P_{t-1} A^\top \\
        \mu_{t|t-1} & = A \mu_{t-1} + \mu_w
    \end{align}
    \end{subequations}
    \item \emph{Measurement update:}
    \begin{subequations}\label{eq: KF meas update}
    \begin{align}
        P_{t}^{-1} & = P_{t|t-1}^{-1} + C^\top \Sigma_v^{-1}C \\
        \mu_{t} & = \mu_{t|t-1} + P_{t} C^\top \Sigma_v^{-1} (y_t - C\mu_{t|t-1} - \mu_v)
    \end{align}
    \end{subequations}
\end{itemize}

\subsection{The Fenchel Conjugate}
The Fenchel conjugate of a function, also known as the convex conjugate or simply conjugate, is defined as follows.
\begin{definition}\label{def: fenchel}
    For a function $g:\R^n \to \R\cup\{\infty\}$, the conjugate function $g^*$ is defined as \cite[\S 3.3]{boyd2004convex}
    \begin{equation*}
        g^*(\lambda) = \sup_{x} \big( \lambda^\top x - g(x) \big).
    \end{equation*}
    Some useful properties: the conjugate $g^*$ is a always a convex function, and if $g$ itself is convex, then $g^{**} \defeq (g^*)^* = g$.
\end{definition}

\begin{proposition}\label{prop: Fenchel of general quadratic}
Given a matrix $M \succ 0$, vector $m$ and scalar $\gamma$ of compatible dimensions, the conjugate of a convex quadratic function is given as follows.
\begin{align*}
\text{If }
g(x) &= \frac{1}{2}\bmat{x\\1}^\top
\bmat{M & m\\ m^\top &\gamma}
\bmat{x\\1},\; \text{then} \\
g^*(\lambda) &= \frac{1}{2}\bmat{\lambda\\1}^\top
\bmat{M^{-1} & -M^{-1}m\\-m^\top M^{-1} &m^\top M^{-1} m-\gamma}
\bmat{\lambda\\1}.
\end{align*}
\end{proposition}
\smallskip

\subsection{Dynamic Programming for MAP Estimation}

It was recently observed by Lange \cite{lange2024bellman} that MAP estimators satisfy a Bellman-style recursion different from the Bayesian recursion \eqref{eq:bayesian_update}. We briefly review this result.
Consider the dynamical system \eqref{eq: general dyn sys}.  
The MAP estimator for $x_t$ given measurements $y_0,\dots,y_t$ is found by maximizing the log-posterior $\log{\prob(x_{0:t}\mid y_{0:t})}$, which is equivalent to maximizing the joint log-likelihood of the states and measurements, denoted $L(x_{0:t},y_{0:t})$.\footnote{This is the \emph{trajectory} or \emph{batch} MAP, which maximizes the conditional distribution of the full state history. Some authors also consider the \emph{pointwise} or \emph{filtering} MAP, which instead maximizes the conditional distribution of the current state. One way to compute this pointwise MAP estimate is to use the Bayesian update \eqref{eq:bayesian_update} but replace the expectation in \eqref{eq: mmse estimator} with an $\argmax$.}
By the probability chain rule\looseness=-1
\begin{equation}\label{eq: joint log lklhd}
    L(x_{0:t},y_{0:t})
    = \ell(x_0) + \sum_{i=1}^{t}\ell(x_i \mid x_{i-1}) + \sum_{i=0}^{t}\ell(y_i\mid x_i).
\end{equation}
\begin{definition}
    The \emph{value function} $V_t:\R^n\to\R$ is the joint negative log-likelihood minimized over past states
    \begin{equation} \label{eq: value function def}
        V_t(x) = \min_{x_0, \dots, x_{t-1}} -L(x_{0:t-1}, x, y_{0:t}).
    \end{equation}
    \vspace{-2mm}
\end{definition}

\noindent Using \eqref{eq: joint log lklhd} and \eqref{eq: value function def} we can obtain a Bellman-style recursion.
\smallskip

\begin{lemma}[Bellman recursion]\label{lem: Bellman}
    The value function given in \eqref{eq: value function def} satisfies the following forward recursion 
\begin{equation}\label{eq: bellman recursion}
    V_{t}(x) = -\ell(y_t\mid x) + \min_{\xi}\big\{-\ell(x\mid \xi) + V_{t-1}(\xi)\big\}.
\end{equation}
\end{lemma}
\smallskip

In the Gaussian setting (see \cref{subsec: KF}) the value function is the convex quadratic
\begin{equation}\label{eq: approximate value function}
    V_t(x) = \frac{1}{2}\bmat{x\\1}^\top
    \bmat{P_t^{-1} & -P_t^{-1}\mu_t\\ \star & \dots}
    \bmat{x\\1},
\end{equation}
where $\mu_t$ and $P_t \succ 0$ are the mean and the covariance of the posterior at time $t$. In this setting, the Bellman update \eqref{eq: bellman recursion}, just like the Bayesian update \eqref{eq:bayesian_update}, reduces to the standard KF.
\smallskip

\section{Main Results} \label{sec: main results}

Our starting point is the Bellman recursion of \cref{lem: Bellman}.

\subsection{Quadratic approximation}\label{sec: quadratic approximation}

In this paper, we approximate the value function \eqref{eq: value function def} with a quadratic function of the form \eqref{eq: approximate value function}. The justification for this approximation is that many log-PDFs resemble quadratic functions near their modes \cite{lange2024bellman}. Quadratic value functions also reduce Bellman recursions to a pair of algebraic equations via the Fenchel dual, a fact observed in \cite{hassibi2025beyond}.

\begin{lemma}\label{lemma: main result}
    Consider the linear system \eqref{eq: general dyn sys} with $q(\cdot)$ and $r(\cdot)$ defined in \cref{assumptions}. If $V_t$ and $V_{t-1}$ are quadratic functions of the form \eqref{eq: approximate value function}, then there exists a function $p(\cdot)$ such that the Bellman recursion \eqref{eq: bellman recursion} simplifies to the following pair of algebraic equations
    \begin{subequations}\label{eq: the two eqs}
    \begin{align}
    &p(x) + r(y_t - Cx) = \frac{1}{2}
    \bmat{x \\ 1}^\top \!\!
    \squeezemat\bmat{P_{t}^{-1} & -P_{t}^{-1}\mu_{t}\\\star & \dots}\!
    \bmat{x\\1}, \label{eq:first} \\
    &p^*(\lambda) - q^*(\lambda) =  \frac{1}{2}
    \squeezemat\bmat{\lambda \\ 1}^\top \!\!
    \bmat{AP_{t-1}A^\top & A\mu_{t-1}\\ \star & \dots}\!
    \bmat{\lambda\\1}\!.\label{eq:second}
    \end{align}
    \end{subequations}
\end{lemma}
\smallskip

\begin{proof}
See \cref{app: proof of lemma}.
\end{proof}
\medskip

\begin{corollary}\label{corollary 1}
    In the setting of \cref{lemma: main result}, if we further assume the process noise is Gaussian with $w_t \sim \N(\mu_w, \Sigma_w)$, then \cref{eq: the two eqs} simplifies to the single equation
    \begin{multline}\label{eq: the one equation}
    \frac{1}{2}\bmat{x\\1}^\top\!
    \squeezemat\bmat{P_t^{-1} & -P_t^{-1}\mu_t\\ \star & \dots}\!
    \bmat{x\\1}\\
    =  \frac{1}{2}\bmat{x\\1}^\top\!
    \squeezemat\bmat{P_{t \mid t-1}^{-1} & -P_{t|t-1}^{-1}\mu_{t|t-1} \\ \star & \dots}\!
    \bmat{x\\1}+ r(y_t - Cx),
    \end{multline}%
where $P_{t|t-1}$ and $\mu_{t|t-1}$ are given in \eqref{eq: KF time update}.
\end{corollary}
\smallskip
\begin{proof}
See \cref{app: proof of corr}.
\end{proof}

\begin{figure*}
    \centering
    \includegraphics[width=\textwidth]{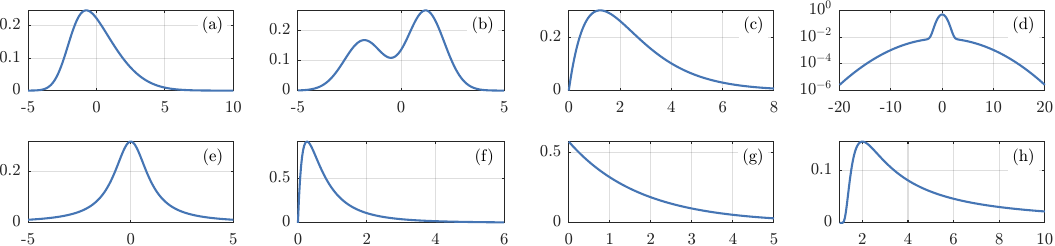}
    \caption{PDFs of measurement noise distributions used in our numerical experiments:
    \textbf{(a)}~Skewed normal,
    \textbf{(b)}~Bimodal Gaussian mixture,
    \textbf{(c)}~Gamma,
    \textbf{(d)}~Impulsive Gaussian mixture (on a log scale),
    \textbf{(e)}~Cauchy,
    \textbf{(f)}~Beta prime,
    \textbf{(g)}~Exponential, and
    \textbf{(h)}~L\'evy
    distributions.
    Refer to \cref{tab: distributions} for exact PDFs.}  
    \label{fig:PDF samples}
\end{figure*}

\subsection{Proposed Estimator}\label{sec: proposed estimators}

For the rest of this paper, we assume the process noise is Gaussian, i.e., $q$ is quadratic, and use \cref{corollary 1} to derive our estimator. The more general case of non-Gaussian process noise is left for future studies. 

Our proposed estimator is based on using the quadratic approximation for the value function described in \cref{sec: quadratic approximation}, together with a specialized quadratic approximation of $r$.

We propose a quadratic approximation for $r$ \emph{about a given point $\bar v$} in the following sense. We choose the Hessian $M_r$ of the approximation by fitting a quadratic whose gradient matches that of $r$ at $\bar v$ and whose vertex is at the \emph{minimizer of $r$} (the mode $m_v$ of the measurement noise distribution). This leads to the equation $M_r (\bar v - m_v) = \grad r(\bar v)$. Among possible choices of $M_r$, we opt for the simple diagonal solution\looseness=-1
\begin{equation}\label{Mg}
M_r  = \diag\biggl(\frac{\left[\grad{r}(\bar v)\right]_i}{\bar v_i - [m_v]_i}\biggr),
\end{equation}
where $[x]_i$ denotes the $i\textsuperscript{th}$ component of the vector $x$.

We require $M_r \succ 0$, which will typically be true, but may not be the case if the distribution is bimodal or has finite support. For such cases, we make some sensible adjustments, which we detail in \cref{rem:special1,rem:special2}.
The complete quadratic approximation therefore has the form
\begin{equation}\label{eq:g}
r(v) \approx \frac{1}{2}(v-m_v)^\top M_r (v-m_v) + \text{(constant)},
\end{equation}
with $M_r$ defined in \eqref{Mg}.
To approximate $r(v)$ in \eqref{eq: the one equation} we use $\bar v \coloneq y_t-C\mu_{t\mid t-1}$, yielding the approximation
\begin{align*}
    r(y_t-Cx) & \approx
    \frac{1}{2}
    \bmat{x\\1}^\top\!
    \squeezemat\bmat{C^\top M_rC & -C^\top M_r (y_t - m_v)\\ \star & \dots}\!
    \bmat{x\\1}.
\end{align*}
So, rather than approximating $r(v)$ globally as in the standard KF, we use an approximation that is more accurate \emph{near our current estimate of $v$}.
Substituting the approximation of $r(y_t-Cx)$ into \eqref{eq: the one equation}  and simplifying yields our proposed estimator, given below.
\smallskip

\textbf{Proposed estimator.} Given prior parameters $\mu_{t-1}$, $P_{t-1}$ and new measurement $y_t$, we find $\mu_t$, $P_t$ via the updates
\begin{subequations}\label{eq: proposed estimator gaussian process}
\begin{align}
P_{t|t-1} & = \Sigma_w + A P_{t-1} A^\top \label{qq1}\\
        \mu_{t|t-1} & = A \mu_{t-1} + \mu_w \label{qq2}\\
    P_t^{-1} & = P_{t|t-1}^{-1} + C^\top M_rC, \label{qq3}\\
    \mu_t & = \mu_{t|t-1} + P_tC^\top \grad{r}(\bar v), \label{qq4}
\end{align}
\end{subequations}
with $\bar v = y_t - C\mu_{t|t-1}$ and $M_r$ given in \eqref{Mg}.
\smallskip

\begin{remark}\label{remark: reduce to kf}
    Since we assume Gaussian process noise, \cref{qq1,qq2} are the same as the time update \eqref{eq: KF time update} of the standard KF.
    When the measurement noise is Gaussian with a diagonal covariance matrix, \cref{qq3,qq4} reduce to the measurement update \eqref{eq: KF meas update} of the standard KF.
\end{remark}

\begin{remark}\label{rem:newton}
    Consider the optimization problem
    \begin{equation}\label{eq: explicit as optimization}
        \min_{\xi} \biggl\{\frac{1}{2} (\xi - \mu_{t \mid t-1})^\top P_{t \mid t-1}^{-1} (\xi - \mu_{t \mid t-1}) + r(y_t - C\xi)\biggr\},
    \end{equation}
    where the objective is the negative log-posterior distribution $\ell(x_t = \xi \mid y_{0:t})=\ell(x_t = \xi \mid y_{0:t-1}) + \ell(y_t \mid x_t = \xi)$. For Gaussian process noise, the prior $\prob(x_t = \xi \mid y_{0:t-1})$ is Gaussian. Using our quadratic approximation for $r(y_t-C\xi)$ about the point $\xi = \mu_{t|t-1}$ and substituting into \eqref{eq: explicit as optimization}, and solving for $\xi$ yields a \emph{Newton-like} update that precisely recovers our estimator \eqref{eq: proposed estimator gaussian process}. One can envision other possible estimators that perform multiple Newton-like steps per measurement, or use different iterative schemes altogether.
\end{remark}

\begin{remark}
    Since our proposed approach is similar to a Newton step (see \cref{rem:newton}), one may be tempted to use the second-order Taylor approximation $M_r=\hess r(\bar v)$ and obtain an exact Newton step. However, Newton's method is prone to numerical issues when the Hessian is indefinite or nearly singular (as in bimodal or heavy-tailed distributions). Anchoring the quadratic approximation to the mode of the distribution as in \eqref{eq:g} gives our estimator added robustness.
\end{remark}

\section{Discussion}\label{sec: discussion}

Our estimator \eqref{eq: proposed estimator gaussian process} has similar update equations to those of the Masreliez estimator \cite{masreliez1975approximate}, which uses a score function. However, the Masreliez estimator is fundamentally different because it approximates the Bayesian update \eqref{eq:bayesian_update}, while our approach uses the Bellman update \eqref{eq: bellman recursion}. The Masreliez estimator directly approximates the posterior distribution as a Gaussian, while we approximate the \emph{noise distribution} at each iteration based on the most recent state estimate. As a result, our estimator is comparable to the standard KF in terms of computational footprint, while the Masreliez estimator is significantly more expensive. Specifically, in the Masreliez filter, in order to compute the score function at each timestep, the density of the predicted measurement needs to be computed via convolution, i.e.,
\begin{equation}\label{eq: mas conv}
    \prob(y_t \mid y_{0:t-1}) =
    \int \prob(y_t \mid x_t)\,\prob(x_t \mid y_{0:t-1}) \dd x_t,
\end{equation}
where the prior $\prob(x_t \mid y_{0:t-1})$ is assumed to be Gaussian. A common method to compute integrals of the form \eqref{eq: mas conv} is Gauss--Hermite quadrature, which has the computation complexity of $O(s^n)$, where $s$ is the number of points used to grid the integration domain. As a result, the total complexity of the Masreliez estimator is $O(n^3+s^n)$, while the complexity of our method is $O(n^3)$, which is the same as the standard KF. 
As we will see in \cref{sec: numerical examples}, our estimator has comparable performance to the Masreliez filter in terms of RMSE, while requiring significantly less computation. 

For a linear dynamical system with non-Gaussian noise, the Kalman filter is the \emph{best linear unbiased estimator} (BLUE) \cite[\S 5.2]{simon2006optimal}. Our proposed estimator \eqref{eq: proposed estimator gaussian process} is different from a standard KF where noise is \emph{globally} approximated by a Gaussian. Rather, we use \emph{local} approximations of the noise distribution at each timestep, which leads to a nonlinear estimator and improved empirical performance compared to the standard KF (see \cref{sec: numerical examples}).

\section{Numerical Examples} \label{sec: numerical examples}
We evaluated the performance of our proposed estimator \eqref{eq: proposed estimator gaussian process} on the following 2D rotational system \cite{chen2017maximum}
\begin{subequations}\label{eq: dyn sys example1}
\begin{align}
    \bmat{x_{1,t}\\x_{2,t}} & =
    \squeezemat\bmat{\cos\frac{\pi}{18} & -\sin\frac{\pi}{18} \\ \sin\frac{\pi}{18} &\hphantom{-}\cos\frac{\pi}{18}}\!
    \bmat{x_{1,t-1}\\x_{2,t-1}} + \bmat{w_{1,t}\\w_{2,t}}, \\
    y_t & = \bmat{1 & 1} x_t + v_t.
\end{align}
\end{subequations}
For the process noise, we used $w_t \sim \N(0, 0.05I_2)$. For the measurement noise, we used $v_t \sim \N(0,3)$ as our baseline. We then tested a variety of measurement noise distributions (see \cref{fig:PDF samples,tab: distributions}). To ensure consistency across experiments, our noise PDFs were adjusted so that their first and second moments, matched those of the baseline measurement noise whenever possible.

\begin{table*}[t]
  \centering
  \caption{Various noise distributions used in numerical experiments. All distributions are configured to share a common mean and variance whenever possible.  In this table, $\phi(\cdot)$ and $\Phi(\cdot)$ represent the PDF and cumulative distribution function (CDF) of a standard Gaussian distribution, respectively. Furthermore, $\Gamma(\cdot)$ and $B(\cdot, \cdot)$ denote the gamma and beta functions, respectively. These PDFs are plotted in \cref{fig:PDF samples}.}
  \label{tab:noise}
  \begin{tabularx}{\linewidth}{lYYYY}
    \toprule
    Noise type &
    Skew normal (a) &
    Bimodal (b) &
    Gamma (c) &
    Impulsive (d) \\
    \midrule
    Noise PDF &
    $\frac{2}{\omega}\phi\Bigl(\tfrac{x-\xi}{\omega}\Bigr)
    \Phi\Bigl(\alpha\,\frac{x-\xi}{\omega}\Bigr)$ &
    $\sum_{i=1}^{2}\alpha_i \,\phi\Bigl(\frac{x-\mu_i}{\sigma_i}\Bigr)$ &
    $\dfrac{1}{\Gamma(\alpha)\theta^\alpha}\,x^{\alpha-1}\exp\Bigl(\frac{-x}{\theta}\Bigr)$ &
    $\sum_{i=1}^{2}\alpha_i \,\phi\Bigl(\frac{x-\mu_i}{\sigma_i}\Bigr)$ \\
    \addlinespace
    Parameters &
    $\begin{aligned}
        \xi &= -2.0063 \\[-2pt]
        \omega &= 2.6505 \\[-2pt]
        \alpha &= 3
    \end{aligned}$ &
    $\begin{aligned}
        \alpha_1 &= 0.4 & \alpha_2 &= 0.6 \\[-2pt]
        \mu_1 &= -1.8    & \mu_2 &= 1.2  \\[-2pt]
        \sigma_1^2 &= 0.9 & \sigma_2^2 &= 0.8
    \end{aligned}$ &
    $\begin{aligned}
        \alpha &= 2 \\[-2pt]
        \theta &= \sqrt{3/2}
    \end{aligned}$ &
    $\begin{aligned}
        \alpha_1 &= 0.1 & \alpha_2 &= 0.9 \\[-2pt]
        \mu_1 &= 0 & \mu_2 &= 0 \\[-2pt]
        \sigma_1^2 &= 25 & \sigma_2^2 &= 0.5556
    \end{aligned}$ \\
    \addlinespace
    \toprule
    Noise type &
    Cauchy (e) &
    Beta prime (f) &
    Exponential (g) &
    Lévy (h) \\
    \midrule
    Noise PDF &
    $\dfrac{1}{\pi}\,\dfrac{\gamma}{(x-x_0)^2+\gamma^2}$ &
    $\dfrac{x^{\alpha-1}(1+x)^{-\alpha-\beta}}{B(\alpha,\beta)}$ &
    $\lambda \exp(-\lambda x)$ &
    $\frac{\sqrt{c / (2\pi)}}{(x-\mu)^{3/2}} \exp\Bigl(\tfrac{-c}{2(x-\mu)}\Bigr)$ \\
    \addlinespace
    Parameters &
    $x_0=0\qquad \gamma=1$ &
    $\alpha=2 \qquad \beta=2.7891$ &
    $\lambda=1/\sqrt{3}$ &
    $\mu=1\qquad c=3$ \\
    \bottomrule
  \end{tabularx}
  \label{tab: distributions}
\end{table*}
For each noise distribution, we performed $200$ Monte Carlo runs with different noise realizations, each time simulating \eqref{eq: dyn sys example1} for $200$ timesteps and reporting the mean and standard deviation of the final RMSE and the geometric mean of computation time. We compared our proposed estimator to the standard KF, the Masreliez filter \cite{masreliez1975approximate} and the MCKF \cite{chen2017maximum}. We also ran a particle filter (PF), which served as a minimum MSE baseline. For methods with tuning parameters, we tried different tunings and chose the most performant one. Specifically, for the PF we used $1000$ particles and for the MCKF we used a kernel size of $5$. For the Masreliez filter, the integral in \eqref{eq: mas conv} was evaluated numerically over a domain spanning $\pm5$ standard deviations from the prior estimate, discretized into a grid of $50$ points. Estimators were implemented in MATLAB and run on a conventional laptop.

Simulation results are shown in \cref{tab:experiments_explicit}. We observe that our estimator outperforms the standard KF under all measurement noise distributions, while having comparable computation time. The Masreliez filter performs slightly better than our estimator, however its computation time is significantly higher than that of our proposed estimator.

With the exception of the standard KF, all the filters we tested were nonlinear and provided little in the way of theoretical guarantees, such as stability of the error dynamics. Indeed, the Masreliez filter was found to diverge in certain cases with heavy-tailed measurement noise.

\begin{figure}[htb]
    \centering
    \includegraphics{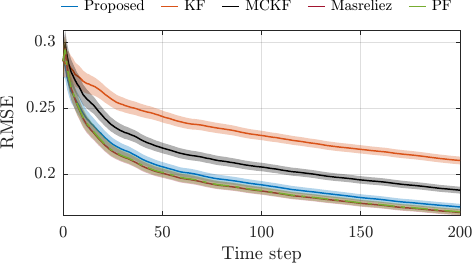}
    \caption{Performance of the proposed estimator \eqref{eq: proposed estimator gaussian process} compared with a variety of estimators under Gaussian process noise and impulsive measurement noise (Item (d) in \cref{fig:PDF samples,tab:noise}). We plot the mean RMSE of $1000$ trials. The shaded area shows the 95\% confidence interval for the mean.}
    \label{fig: rmse beta}
    \vspace{-2mm}
\end{figure}

\noindent A typical RMSE plot is shown in \cref{fig: rmse beta}.
\smallskip

\begin{remark}\label{rem:special1}
    Special consideration is required when computing $M_r$ for the bimodal distribution (b), as the \cref{Mg} assumes a single mode. To resolve this ambiguity, we use the direction of the gradient at $\bar{v}$. When $\bar v$ is between the two modes, for each $i$, we choose $m_v$ to be the mode such that $[\grad r(\bar v)]_i/(\bar v_i-[m_v]_i) > 0$, thus ensuring $M_r \succ 0$.
\end{remark}
\begin{remark}\label{rem:special2}
    For distributions with compact support, whenever $\bar v$ falls outside of the valid domain, we perform a projection with $\epsilon$-margin. For example, the support of the L\'evy distribution is $(\mu,\infty)$. Thus, we use $\bar v = \max(\bar v, \mu + \epsilon)$, where $\epsilon$ is a vector of small positive entries.
\end{remark}

Code to generate all figures and tables can be found here: \url{https://github.com/QCGroup/dpkf}.

\begin{table*}
\addtolength{\tabcolsep}{4pt}
\caption{Simulation results for the system \eqref{eq: dyn sys example1} with Gaussian process noise and various measurement noise distributions (see \cref{fig:PDF samples,tab: distributions}). We report the mean and standard deviation of the RMSE over $200$ Monte Carlo runs, and the geometric mean runtime, normalized so that KF $=1.00$.}\label{tab:experiments_explicit}
\centering
\begin{tabular}{l cr cr cr cr}
    \toprule
    & \multicolumn{2}{c}{Skewed normal (a)} & \multicolumn{2}{c}{Bimodal (b)} & \multicolumn{2}{c}{Gamma (c)} & \multicolumn{2}{c}{Impulsive (d)} \\
    \cmidrule(lr){2-3} \cmidrule(lr){4-5} \cmidrule(lr){6-7} \cmidrule(lr){8-9}
    Estimator & RMSE $\pm$ s.d. & Time & RMSE $\pm$ s.d. & Time  & RMSE $\pm$ s.d. & Time  & RMSE $\pm$ s.d. & Time \\
    \midrule
    KF & $0.213 \pm 0.050$ & $1.00$ & $0.217 \pm 0.051$ & $1.00$ & $0.213 \pm 0.045$ & $1.00$ & $0.212 \pm 0.048$ & $1.00$\\
    MCKF  & $0.212 \pm 0.050$ & $6.45$ & $0.318 \pm 0.064$ & $6.63$ & $0.212 \pm 0.045$ & $6.48$ & $0.191 \pm 0.043$  & $8.20$\\ 
    Masreliez & $0.206 \pm 0.049$ & $56.48$ & $0.202 \pm 0.045$  & $39.87$ & diverged & --- & $0.173 \pm 0.038$ & $37.28$\\
    Ours \eqref{eq: proposed estimator gaussian process} & $0.205 \pm 0.049$ & $0.93$ & $0.209 \pm 0.048$ & $1.47$ & $0.198 \pm 0.043$ & $0.82$ & $0.177 \pm 0.038$  & $1.34$\\
    PF & $0.205 \pm 0.049$ & $63.67$ & $0.201 \pm 0.045$ & $65.27$ & $0.191 \pm 0.041$ & $76.02$ & $0.173 \pm 0.038$ & $62.95$\\
    \addlinespace
    \toprule
    & \multicolumn{2}{c}{Cauchy (e)} & \multicolumn{2}{c}{Beta prime (f)} & \multicolumn{2}{c}{Exponential (g)} & \multicolumn{2}{c}{L\'{e}vy (h)} \\
    \cmidrule(lr){2-3} \cmidrule(lr){4-5} \cmidrule(lr){6-7} \cmidrule(lr){8-9}
    Estimator & RMSE $\pm$ s.d. & Time & RMSE $\pm$ s.d. & Time  & RMSE $\pm$ s.d. & Time  & RMSE $\pm$ s.d. & Time \\
    \midrule
    KF & N/A & --- & $0.204 \pm 0.048$  & $1.00$ & $0.212 \pm 0.049$  & $1.00$ & N/A & ---\\
    MCKF$^{\dagger}$
    & N/A & --- & $0.201 \pm 0.047$ & $6.82$ & $0.206 \pm 0.046$  & $8.90$ & N/A  & ---\\ 
    Masreliez & $0.216 \pm 0.047$  & $28.49$ & diverged & --- & diverged & --- & diverged & ---\\
    Ours \eqref{eq: proposed estimator gaussian process} & $0.228 \pm 0.052$  & $1.00$ & $0.184 \pm 0.043$ & $0.92$ & $0.203 \pm 0.050$  & $0.90$ & $0.254 \pm 0.058$ & $1.00$\\
    PF & $0.216 \pm 0.048$  & $51.71$ & $0.156 \pm 0.034$ & $80.25$ & $0.177 \pm 0.040$  & $77.73$ & $0.242 \pm 0.053$  & $76.88$\\
    \bottomrule
    \end{tabular}
    \begin{minipage}{\textwidth}
    \footnotesize
    \vspace{2pt}
    $^{\dagger}$ MCKF’s derivation assumes finite noise covariance, which is undefined for Cauchy and L\'{e}vy distributions. In practice, one can still use it in these cases, but it requires manually tuning a nominal covariance parameter.
    \end{minipage}
\end{table*}

\section{Conclusion and Future Work}\label{sec: conclusion and future works}
In this paper, we studied MAP estimation for linear systems under non-Gaussian noise from an optimization perspective. We showed that while the general Bellman recursion for MAP estimation does not have a closed-form solution, it can be reduced to algebraic equations via approximation of the value function. To locally approximate noise distributions, we introduced a novel robust quadratic approximation. These results were then leveraged to design a nonlinear state estimator under Gaussian process and non-Gaussian measurement noise. Simulations demonstrated that our proposed estimator outperformed the standard KF and had comparable performance to state-of-the-art estimators, highlighting its practical effectiveness. Moreover, our proposed estimator is computationally efficient, requiring only modest overhead compared to the standard Kalman Filter.

The estimator designed in this paper was for systems with Gaussian process and non-Gaussian measurement noise. One immediate extension is to use \cref{lemma: main result} to derive estimators for cases in which the process noise is also non-Gaussian. Another future direction is generalizing \cref{lemma: main result} to derive smoothers and predictors based on the Bellman recursion \eqref{eq: bellman recursion}, or to generalize our work to nonlinear dynamical systems and develop analogs of the extended or unscented Kalman filter.
Finally, more extensive numerical and theoretical analyses are needed to better characterize the performance and limitations of our proposed estimator. 

\appendix

\subsection{\texorpdfstring{Proof of \cref{lemma: main result}}{Proof of Lemma~\ref{lemma: main result}}}
\label{app: proof of lemma}

Substituting the approximation \eqref{eq: approximate value function} for $V_t$ and $V_{t-1}$ and 
$\ell(y_t\mid x) = -r(y_t - Cx)$
and
$\ell(x\mid \xi) = -q(x-A\xi)$
into the Bellman recursion \eqref{eq: bellman recursion}, we obtain 
\begin{align}\label{eq: proof step 1}
& \frac{1}{2}\bmat{x\\1}^\top
    \bmat{P_t^{-1} & -P_t^{-1}\mu_t\\ \star & \dots}
    \bmat{x\\1} = r(y_t - Cx) \notag \\ 
& + \min_\xi\Bigg\{ q(x - A\xi)
    +\frac{1}{2}\bmat{\xi\\1}^\top
    \bmat{P_{t-1}^{-1} & -P_{t-1}^{-1}\mu_{t-1}\\ \star & \dots}
    \bmat{\xi\\1} \Bigg\}.
\end{align}
We then define 
\begin{equation*}\label{eq: proof step 3}
    p(x) \defeq \frac{1}{2}\bmat{x\\1}^\top\!
    \squeezemat\bmat{P_t^{-1} & -P_t^{-1}\mu_t\\\star & \dots}\!
    \bmat{x\\1} - r(y_t - Cx),
\end{equation*}
which rearranges to \eqref{eq:first}.
Expressing \eqref{eq: proof step 1} in terms of $p(x)$,
\begin{equation*}
    p(x) = \min_\xi\Bigg\{ q(x - A\xi) 
    +\frac{1}{2}\!\bmat{\xi\\1}^\top\!\!
    \squeezemat\bmat{P_{t-1}^{-1} & -P_{t-1}^{-1}\mu_{t-1}\\\star & \dots}\!\!
    \bmat{\xi\\1} \!\Bigg\}.
\end{equation*}
Taking the conjugate of both sides via \cref{def: fenchel} and simplifying, we obtain \eqref{eq:second}, which completes the proof.\hfill\QED

\subsection{\texorpdfstring{Proof of \cref{corollary 1}}{Proof of Corollary~\ref{corollary 1}}}\label{app: proof of corr}

Since $w \sim \N(\mu_w, \Sigma_w)$ we have
\begin{align}\label{eq: q for Gaussian}
    q(w) & = \frac{1}{2}(w-\mu_w)^\top\Sigma_w^{-1}(w-\mu_w) + \text{(constant)}.
\end{align}
Taking the conjugate of \eqref{eq: q for Gaussian} using \cref{prop: Fenchel of general quadratic} we get
\begin{equation}\label{eq: fenchel of q(w)}
    q^*(\lambda) = \frac{1}{2}\begin{bmatrix}
    \lambda\\1
\end{bmatrix}^\top
\begin{bmatrix}
    \Sigma_w & \mu_w\\
    \star & \dots  
\end{bmatrix}
\begin{bmatrix}
    \lambda \\ 1
\end{bmatrix}.
\end{equation}
Substitute \eqref{eq: fenchel of q(w)} into \eqref{eq:second} and use \eqref{eq: KF time update}
to obtain 
\begin{equation}\label{eq: p^*(lambda) Gaussian}
    p^*(\lambda) =  \frac{1}{2}\begin{bmatrix}
    \lambda \\ 1
\end{bmatrix}^\top\begin{bmatrix}
     P_{t|t-1} & \mu_{t|t-1}\\
    \star & \dots
\end{bmatrix}\begin{bmatrix}
    \lambda \\ 1
\end{bmatrix}.
\end{equation}
Taking the conjugate of \eqref{eq: p^*(lambda) Gaussian} using \cref{prop: Fenchel of general quadratic} yields
    \begin{equation}\label{eq: p(x) Gaussian from q}
    p(x) = \frac{1}{2}\begin{bmatrix}
        x\\1
    \end{bmatrix}^\top
    \begin{bmatrix}
        P_{t \mid t-1}^{-1} & -P_{t|t-1}^{-1}\mu_{t|t-1} \\
        \star & \dots
    \end{bmatrix}
    \begin{bmatrix}
        x\\1
    \end{bmatrix}.
\end{equation}
Substituting \eqref{eq: p(x) Gaussian from q} into \eqref{eq:first} yields \eqref{eq: the one equation}, as required.\hfill\QED
\bigskip

\bibliographystyle{IEEEtran}
\bibliography{references}

\begin{thebibliography}{10}
\providecommand{\url}[1]{#1}
\csname url@samestyle\endcsname
\providecommand{\newblock}{\relax}
\providecommand{\bibinfo}[2]{#2}
\providecommand{\BIBentrySTDinterwordspacing}{\spaceskip=0pt\relax}
\providecommand{\BIBentryALTinterwordstretchfactor}{4}
\providecommand{\BIBentryALTinterwordspacing}{\spaceskip=\fontdimen2\font plus
\BIBentryALTinterwordstretchfactor\fontdimen3\font minus
  \fontdimen4\font\relax}
\providecommand{\BIBforeignlanguage}[2]{{%
\expandafter\ifx\csname l@#1\endcsname\relax
\typeout{** WARNING: IEEEtran.bst: No hyphenation pattern has been}%
\typeout{** loaded for the language `#1'. Using the pattern for}%
\typeout{** the default language instead.}%
\else
\language=\csname l@#1\endcsname
\fi
#2}}
\providecommand{\BIBdecl}{\relax}
\BIBdecl

\bibitem{han2020blind}
Q.~Han, L.~Yang, J.~Du, and L.~Cheng, ``Blind equalization for chaotic signals
  based on echo state network and {K}alman filter under nonlinear channels,''
  \emph{IEEE Communications Letters}, vol.~25, no.~2, pp. 589--592, 2020.

\bibitem{abdollahi2024improved}
M.~Abdollahi, S.~H. Pourtakdoust, M.~Y. Nooshabadi, and H.~N. Pishkenari, ``An
  improved multi-state constraint {K}alman filter for visual-inertial
  odometry,'' \emph{Journal of the Franklin Institute}, vol. 361, no.~15, p.
  107130, 2024.

\bibitem{taassori2024enhancing}
M.~Taassori and B.~Vizv{\'a}ri, ``Enhancing medical image denoising: A hybrid
  approach incorporating adaptive {K}alman filter and non-local means with
  {L}atin square optimization,'' \emph{Electronics}, vol.~13, no.~13, p. 2640,
  2024.

\bibitem{wang2020novel}
J.~Wang, J.~Li, S.~Yan, W.~Shi, X.~Yang, Y.~Guo, and T.~A. Gulliver, ``A novel
  underwater acoustic signal denoising algorithm for {G}aussian/non-{G}aussian
  impulsive noise,'' \emph{IEEE Transactions on Vehicular Technology}, vol.~70,
  no.~1, pp. 429--445, 2020.

\bibitem{sarfi2019decentralized}
V.~Sarfi, A.~Ghasemkhani, I.~Niazazari, H.~Livani, and L.~Yang, ``Decentralized
  dynamic state estimation with bimodal {G}aussian mixture measurement noise,''
  in \emph{2019 North American Power Symposium (NAPS)}.\hskip 1em plus 0.5em
  minus 0.4em\relax IEEE, 2019, pp. 1--5.

\bibitem{st2018automatic}
S.~St-Jean, A.~De~Luca, M.~A. Viergever, and A.~Leemans, ``Automatic, fast and
  robust characterization of noise distributions for diffusion {MRI},'' in
  \emph{International Conference on Medical Image Computing and
  Computer-Assisted Intervention}.\hskip 1em plus 0.5em minus 0.4em\relax
  Springer, 2018, pp. 304--312.

\bibitem{gordon1993novel}
N.~J. Gordon, D.~J. Salmond, and A.~F. Smith, ``Novel approach to
  nonlinear/non-{G}aussian {B}ayesian state estimation,'' in \emph{IEE
  proceedings F (radar and signal processing)}, vol. 140, no.~2.\hskip 1em plus
  0.5em minus 0.4em\relax IET, 1993, pp. 107--113.

\bibitem{elfring2021particle}
J.~Elfring, E.~Torta, and R.~Van De~Molengraft, ``Particle filters: A hands-on
  tutorial,'' \emph{Sensors}, vol.~21, no.~2, p. 438, 2021.

\bibitem{sorenson1971recursive}
H.~W. Sorenson and D.~L. Alspach, ``Recursive {B}ayesian estimation using
  {G}aussian sums,'' \emph{Automatica}, vol.~7, no.~4, pp. 465--479, 1971.

\bibitem{bilik2010mmse}
I.~Bilik and J.~Tabrikian, ``{MMSE}-based filtering in presence of
  non-{G}aussian system and measurement noise,'' \emph{IEEE Transactions on
  Aerospace and Electronic Systems}, vol.~46, no.~3, pp. 1153--1170, 2010.

\bibitem{masreliez1975approximate}
C.~Masreliez, ``Approximate non-{G}aussian filtering with linear state and
  observation relations,'' \emph{IEEE Transactions on Automatic Control},
  vol.~20, no.~1, pp. 107--110, 1975.

\bibitem{wu1989kalman}
W.-R. Wu and A.~Kunda, ``{K}alman filtering in non-{G}aussian environment using
  efficient score function approximation,'' in \emph{1989 IEEE International
  Symposium on Circuits and Systems (ISCAS)}.\hskip 1em plus 0.5em minus
  0.4em\relax IEEE, 1989, pp. 413--416.

\bibitem{chen2017maximum}
B.~Chen, X.~Liu, H.~Zhao, and J.~C. Principe, ``Maximum correntropy {K}alman
  filter,'' \emph{Automatica}, vol.~76, pp. 70--77, 2017.

\bibitem{he2023generalized}
J.~He, G.~Wang, H.~Yu, J.~Liu, and B.~Peng, ``Generalized minimum error entropy
  {K}alman filter for non-{G}aussian noise,'' \emph{ISA transactions}, vol.
  136, pp. 663--675, 2023.

\bibitem{kircher2021optimization}
A.~Kircher, L.~Bako, E.~Blanco, and M.~Benallouch, ``An optimization framework
  for resilient batch estimation in cyber-physical systems,'' \emph{IEEE
  Transactions on Automatic Control}, vol.~67, no.~10, pp. 5246--5261, 2021.

\bibitem{gultekin2017nonlinear}
S.~Gultekin and J.~Paisley, ``Nonlinear {K}alman filtering with divergence
  minimization,'' \emph{IEEE Transactions on Signal Processing}, vol.~65,
  no.~23, pp. 6319--6331, 2017.

\bibitem{simon2006optimal}
D.~Simon, \emph{Optimal state estimation: {K}alman, H infinity, and nonlinear
  approaches}.\hskip 1em plus 0.5em minus 0.4em\relax John Wiley \& Sons, 2006.

\bibitem{boyd2004convex}
S.~P. Boyd and L.~Vandenberghe, \emph{Convex optimization}.\hskip 1em plus
  0.5em minus 0.4em\relax Cambridge university press, 2004.

\bibitem{lange2024bellman}
R.-J. Lange, ``Bellman filtering and smoothing for state--space models,''
  \emph{Journal of Econometrics}, vol. 238, no.~2, p. 105632, 2024.

\bibitem{hassibi2025beyond}
B.~Hassibi, J.~Hajar, and R.~Ghane, ``Beyond quadratic costs in {LQR}: Bregman
  divergence control,'' \emph{arXiv preprint arXiv:2505.00317}, 2025.

\end{thebibliography}
\vfill

\end{document}